\def\beg{\begin{equation}}
\def\eeq{\end{equation}}
\documentstyle[12pt]{article}
\begin{document}
\begin{center}
{\Large{\bf Comments on ``Competition Between Fractional Quantum Hall Liquid
..., by G. Gervais, L. W. Engel, H. L. Stormer, D. C. Tsui, et al
cond-mat/0402169(5 Feb. 2004)".}}
\vskip0.55cm
{\bf Keshav N. Shrivastava}
\vskip0.25cm
{\it School of Physics, University of Hyderabad,\\
Hyderabad  500046, India}
\end{center}
The quantum Hall effect in ultra-high mobility GaAs/AlGaAs
has been measured and plateaus are found at many different fractions.
The resistivity is quantized as $\rho=h/ie^2$ where $i$ exhibits
many different values. The fractions 5/3, 8/5, 11/7, 14/9 and 17/11 fit
the formula $3p\pm 2/(2p\pm 1)$ and it is claimed that $2p$ flux quanta
are attached to the electron.
The fractions 4/11, 7/11, 12/7, 13/8, 15/11 do not fit the expression for
 $i$, even then the authors insist that flux quanta are attached to the
 electron and hence composite fermions are formed.
We report that the interpretation of the experimental data in terms
of CF is incorrect.\\

PACS numbers: 73.43.-f
\vskip1.0cm

\noindent {\bf 1.~ Introduction}

It has been suggested that flux quanta $\phi=hc/e$ exist as independent
particles which can be attached to the electron to form a bound state
consisting of even number of flux quanta and an electron. These composite
fermions (CF) might be searched in the data of quantum Hall effect where
fractionally charged particles may exist. The magnetic field is produced
by currents so that fields and currents both must occur. The theory of
such CF could not be formulated consistently but there is ambition to
discover new theories and new principles.
     In spite of several years of work and drum beating,  the theory of CF
is not yet known. When the theory itself is not known, it is not possible
to see CF in the experimental data. It has been informed many times[1] that ``CF"
does not exist and the experimentalists should refrain from making claims that
they have observed CF but new students keep coming and they read the earlier
papers and say that they have observed the CF. Therefore, it is being asked
in the APS meeting that a suitable ``retraction" should be issued and it should
be announced clearly that CF have not been observed.
     In this comment we point out that the claim of Gervais et al[2] that they
observed any CF is false.

\noindent{\bf 2.~~Gervais et al}

     Gervais et al say that CF has been observed in N=0 and N=1 Landau
levels. The observation of a series $i=3p\pm2/(2p+1)$ is a phenomenology
to churn the numbers. Even if this series is successful in getting the
experimental fraction, it does not mean that "flux is attached to the electron".
The numbers 4/11 and 7/11 could not be obtained from the CF series, even then
the authors insist that there are CF. When every thing has been done to say that
CF are inconsistent, the Columbia-Princeton authors continue to say that there
are CF. When the numbers 5/2 and 7/2 could not be explained, they started saying
that CF are paired. The fractions 5/3, 8/5, 11/7, 14/9, 17/11 fit the expression
$3p\pm2/(2p+1)$ but 12/7, 13/8, 15/11 do not fit. So whether they fit or not
either way the claim is made that CF are observed. Even if the series gives some
fractions correctly it does not prove that fluxex are attached. The fractions
21/5 and 24/5 are also found in the experimental data but now Gervais et al
suggest that incompressible Laughlin[3, 4] liquid may be formed. Now there is no
need of CF but whether Laughlin is relevant to the data or not is yet another
question.
\noindent{\bf3.~~ Conclusions}.
     In conclusion, the interpretation of the experimental data given by
     Gervais et al in terms of CF(flux-attached quasiparticles) is incorrect.
     There are more than 148 fractional charges found in the data,
     so there is no chance for attaching fluxes to electrons without appropriate
     currents. Hence the CF model is wrong.

    The correct interpretation of Stormer's data[5] is given by us[6-17].
\vskip1.0cm

\noindent{\bf4.~~References}
\vskip1.0cm
\begin{enumerate}
\item K. N. Shrivastava, Bull. Am. Phys. Soc. {\bf 48}, 529 (2003)
\item G. Gervais, L. W. Engel, H. L. Stormer, D. C. Tsui et al,cond-mat/0402169.
\item R. B. Laughlin, Phys. Rev. Lett. {\bf 50}, 1395(1983).
\item V. Kalmeyer, and R. B. Laughlin, Phys. Rev. Lett.{\bf 59}, 2095 (1987)
\item H. L. Stormer, Rev. Mod. Phys. {\bf 71}, 875 (1999).
\item K. N. Shrivastava, Phys. Lett. A {\bf 113}, 435 (1986)
\item K. N. Shrivastava, Mod. Phys. Lett.B {\bf 13}, 1087 (1999).
\item K. N. Shrivastava, Mod. Phys. Lett. B {\bf 14}, 1009 (2000).
\item K. N. Shrivastava, in Frontiers of Fundamental Physics 4, edited by B. G. Sidharth and M. V. Altaisky, Kluwer Academic/Plenum NY 2000, pp. 235-249.
\item K. N. Shrivastava, cond-mat/0212552.
\item W. Pan,H. L. Stormer, D. C. Tsui,L. N. Pfeiffer, K. W. Baldwin and K. W. West, Phys. Rev. Lett. {\bf 90}, 016801 (2003).
\item K. N. Shrivastava, cond-mat/0302610
\item K. N. Shrivastava, cond-mat/0303309 and 0303621.
\item K. N. Shrivastava, cond-mat/0304269.
\item S. Syed, M. J. Manfra,Y. J. Wang, H. L. Stormer, and R. J. Molnar,Phys. Rev. B{\bf 67}, 241304(2003)
\item K. N. Shrivastava,cond-mat/0305415.
\item K. N. Shrivastava, Introduction to quantum Hall effect,\\
      Nova Science Pub. Inc., N. Y. (2002).
\end{enumerate}

\end{document}